\documentclass[prb,twocolumn,showpacs,showkeys,amssymb]{revtex4}

\usepackage{graphicx}
\usepackage{dcolumn}
\usepackage{bm}

\begin{document}


\title{Semiquantitative theory of electronic Raman scattering\\ 
 from medium-size quantum dots}
\author{Alain Delgado}
\affiliation{Centro de Aplicaciones Tecnol\'ogicas y
 Desarrollo Nuclear, Calle 30 No 502, Miramar, Ciudad Habana, Cuba}
\email{gran@ceaden.edu.cu}
\author{Augusto Gonzalez}
\affiliation{Instituto de Cibern\'etica, Matem\'atica y F\'{\i}sica, Calle
 E 309, Vedado, Ciudad Habana, Cuba}
\email{agonzale@icmf.inf.cu}
\author{D.J. Lockwood}
\affiliation{Institute for Microstructural Sciences, National Research Council,
 Ottawa, Canada K1A 0R6}
\email{David.Lockwood@nrc-cnrc.gc.ca}

\begin{abstract}
A consistent semiquantitative theoretical analysis of electronic Raman 
scattering from many-electron quantum dots under resonance excitation
conditions has been performed. The theory is based on random-phase-approximation-like
wave functions, with the Coulomb interactions treated exactly, and hole 
valence-band mixing accounted for within the Kohn-Luttinger Hamiltonian 
framework. The widths of intermediate and final states in the scattering 
process, although treated phenomenologically, play a significant role in
the calculations, particularly for well above band gap excitation. The
calculated polarized and unpolarized Raman spectra reveal a great complexity
of features and details when the incident light energy is swept from below,
through, and above the quantum dot band gap. Incoming and outgoing 
resonances dramatically modify the Raman intensities of the single
particle, charge density, and spin density excitations. The theoretical
results are presented in detail and discussed with regard to experimental
observations.
\end{abstract}


\maketitle

\section{Introduction}

The inelastic (Raman) scattering of light by a semiconductor is an
optical process that has proven its usefulness as a spectroscopic
tool to investigate elementary excitations in semiconductors
\cite{ILS1, ILS2, ILS3}. From the theoretical point of view, the
scattering process is described by a simple expression coming from
second-order perturbation theory \cite{text}:

\begin{widetext}
\begin{equation}
A_{fi}\sim \sum_{int} \frac{\langle f,N_i-1,1_f|\hat H^+_{e-r}|int,N_i-1 \rangle
\langle int,N_i-1|\hat H^-_{e-r}|i,N_i \rangle}{h\nu_i-(E_{int}-E_i)+i\Gamma_{int}}.
\label{eq1}
\end{equation}
\end{widetext}

\noindent $A_{fi}$ is the quantum mechanical amplitude for the
transition from the initial (electronic) state, $|i\rangle$, of
energy $E_i$, to the final state, $|f\rangle$. This transition
involves a change in the state of the radiation field. Indeed, the
final state of the electron-photon system, $|f,N_i-1,1_f\rangle$,
contains $N_i-1$ incident photons of energy $h\nu_i$ (one less
than the initial state), and one photon of energy $h\nu_f$ (the
scattered photon). The sum in Eq. (\ref{eq1}) runs over all
intermediate (virtual) states. $\hat H_{e-r}$ is the electron-radiation
interaction Hamiltonian, and $\Gamma_{int}$ is a phenomenological
damping parameter.

From the amplitudes $A_{fi}$, one computes the differential cross
section \cite{text}:

\begin{equation}
\frac{{\rm d}\sigma}{{\rm d}\Omega_f {\rm d}\nu_f}\sim \sum_f
|A_{fi}|^2 \delta (E_i+h\nu_i-E_f-h\nu_f), \label{eq2}
\end{equation}

\noindent where ${\rm d}\Omega_f$ is the element of solid angle
related to the wave vector of the scattered photon. Energy
conservation is expressed by means of the delta function in Eq.
(\ref{eq2}), which is approximated by a Lorentzian:

\begin{equation}
\delta(x-x_f)=\frac{\Gamma_f/\pi}{(x-x_f)^2+\Gamma_f^2}.
\end{equation}

In the present paper, we focus on the Raman scattering in zero
magnetic field from a
quantum dot containing dozens of electrons. Thus, $|i\rangle$ and
$|f\rangle$ are states with $N_e$ electrons. The incident laser
energy, $h\nu_i$, is taken to be resonant with an interband
transition. It means that only the resonant contribution to
$A_{fi}$ is considered in Eq. (\ref{eq1}), \cite{nota} and that
the intermediate states, $|int\rangle$, contain an additional
electron-hole pair.

Eqs. (\ref{eq1}) and (\ref{eq2}) look very simple, but in fact
their evaluation is a cumbersome task because reliable
approximations to the many-particle wave functions $|i\rangle$,
$|int\rangle$ and $|f\rangle$ need to be computed. A widely used
simplified expression is obtained by assuming a constant denominator
in Eq. (\ref{eq1}) and using 
completeness relations for the intermediate and hole states. In
this way, we arrive at the off-resonance approximation
\cite{ILS1}:

\begin{widetext}
\begin{eqnarray}
A_{fi}^{or}\sim &-&\left\langle f\left| \sum_{\alpha,\alpha'}
 \langle\alpha| e^{i(\vec q_i-\vec q_f)\cdot \vec r}|\alpha'\rangle
 \left\{\frac{2}{3} (\vec \varepsilon_i\cdot\vec\varepsilon_f)~
 [\hat e^\dagger_{\alpha\uparrow}\hat e_{\alpha'\uparrow}+
 \hat e^\dagger_{\alpha\downarrow}\hat e_{\alpha'\downarrow} ]
+\frac{i}{3} (\vec \varepsilon_i\times\vec\varepsilon_f)\cdot \hat z~
 [\hat e^\dagger_{\alpha\uparrow}\hat e_{\alpha'\uparrow}-
 \hat e^\dagger_{\alpha\downarrow}\hat e_{\alpha'\downarrow} ]
\right.\right.\right.\nonumber\\
&+& \left.\left.\left.
\frac{i}{3} (\vec \varepsilon_i\times\vec\varepsilon_f)\cdot
(\hat x+i\hat y)~ \hat e^\dagger_{\alpha\uparrow}\hat e_{\alpha'\downarrow}
+\frac{i}{3} (\vec \varepsilon_i\times\vec\varepsilon_f)\cdot
(\hat x-i\hat y)~ \hat e^\dagger_{\alpha\downarrow}\hat e_{\alpha'\uparrow}
    \right\}\right|i\right\rangle,
\label{eq3}
\end{eqnarray}
\end{widetext}

\noindent where $\vec q$
and $\vec\varepsilon$ are the wave vector and the light
polarization vector, respectively, $\alpha$ and $\alpha'$ label
the Hartree-Fock (HF) states for electrons, and $\hat e$ and
$\hat e^{\dagger}$ are electron annihilation and creation operators.
Notice that, in this approximation, the intermediate states play
no role and the Raman amplitude is identified with the structure
functions, i.e., only collective excitations in final states are
supposed to contribute to the Raman peaks. Four terms are
distinguished in Eq. (\ref{eq3}). The first one, proportional to
$\vec \varepsilon_i\cdot\vec\varepsilon_f$, corresponds to
charge-density excitations (CDE). The next three, proportional to
$\vec \varepsilon_i\times\vec\varepsilon_f$, correspond to
spin-density excitations (SDE).

Most of the analysis of Raman experiments in quantum wells
(qwells), wires (qwires) and dots (qdots) are based on expressions
like Eq. (\ref{eq3}) in spite of its known limitations.
Experiments in qwells and qwires under extreme resonance (i.e.,
when the incident laser energy is close to the energy of the
exciton) have revealed Raman peaks associated with single-particle
excitations (SPE) \cite{exp1}. These peaks do not arise from
Eq. (\ref{eq3}) and are known to be related to taking a proper account of
the intermediate (virtual) states \cite{Sarma}. For still higher
excitation energies (i.e., 40-50 meV above the band gap) a
resonant enhancement of Raman intensities for particular values of
$h\nu_i$ have been observed \cite{Danan}. This effect is clearly
not described by Eq. (\ref{eq3}). It has been ascribed to the
existence of incoming and outgoing resonances in the intermediate
states, although the nature of the outgoing resonances is not
completely understood.  The authors of Ref. [\onlinecite{Danan}]
have suggested the presence of higher-order Raman processes to
explain the observed resonances. We shall show that the usual
second-order expression Eq. (\ref{eq1}) with a phenomenological
$\Gamma_{int}$ accounts for these effects.

A review of relevant experimental facts of electronic Raman
scattering in qdots can be found in Ref. [\onlinecite{Lockwood}].
In our opinion, the best experimental results are those reported
in Ref. [\onlinecite{Heitmann}]. As $h\nu_i$ moves from extreme
resonance to 40 meV above it, the observed Raman spectrum evolves
from a SPE-dominated one to a spectrum dominated by collective
excitations. The positions of collective excitations for the dots
studied in Ref. [\onlinecite{Heitmann}] have been computed in Ref.
[\onlinecite{Lipparini}] by means of expressions analogous to Eq.
(\ref{eq3}), but the dependence on $h\nu_i$ could only be obtained
if one starts from Eq. (\ref{eq1}).

In the present paper, we give a consistent theory of Raman
scattering in medium-size qdots (dozens of electrons) based on the
exact expression given in Eq. (\ref{eq1}). The theory is, however,
``semiquantitative'' because random phase approximation (RPA) -
like wave functions and phenomenological $\Gamma_{int}$ and
$\Gamma_f$ are used. The main limitation of the RPA functions in
the present context is not related to the absence of correlation
effects, but to an inadequate description of the density of energy
levels and of the matrix elements of the electron-radiation
interaction Hamiltonian. The main virtue of the RPA functions, on
the other hand, is that collective excitations are described quite
well. In spite of its limitations, the theory is able to reproduce
all of the observed qualitative features of Raman scattering in
qdots.

With respect to previous calculations, we are aware of the exact
computations (i.e., numerically exact electronic wave functions
plus Eq. (\ref{eq1})) for a two-electron quantum ring made in Ref.
[\onlinecite{rusos}], and of the approximate calculations for the
12-electron dot in Ref. [\onlinecite{calc}]. In this paper, we
report calculations for a dot with 42 electrons. Coulomb
interactions are treated exactly (to the extent that the RPA
approximation allows it) both in intermediate and final states.
Valence-band mixing effects for the hole are accounted for in the
framework of the Kohn-Luttinger Hamiltonian.

The plan of the paper is as follows. In the next section, we
derive the theory needed for calculating the resonant Raman
spectra, which are later discussed in section \ref{sec3}.
Conclusions are drawn in section \ref{sec4}.

\section{The theory}
\label{sec2}

The computation of the Raman amplitude $A_{fi}$ requires: (a) the
calculation of HF single-particle states for electrons and holes,
(b) obtaining the final $N_e$-electron states, $|f\rangle$, by
means of the RPA scheme, (c) obtaining the $(N_e+1)$-electron plus
one hole states, $|int\rangle$, by means of the so-called
particle-particle RPA formalism and, finally, (d) the computation
of the matrix elements of the electron-radiation Hamiltonian,
$\hat H_{e-r}$. Additionally, we shall compute matrix elements of
multipole operators, something equivalent to the structure
functions, and the density of final-state energy levels. Many of
the required expressions and formulas were given explicitly in
Ref. [\onlinecite{nuestro}] for the neutral electron-hole system.
They can be used in the present context with minor modifications.

Generally speaking, we use a HF-like scheme to describe the ground
state of the $N_e$-electron system (in fact, the RPA assumes that 
there are some ``correlations'' in the ground state, $|i\rangle$,
as can be seen from the formulas below).
An effective (conduction) mass approximation is used to describe
electrons in the qdot. Thus, the $N_e$-electron problem with
confinement and Coulomb interactions is solved in this way. The
excited states of this system, $|f\rangle$, are looked for with
the help of the RPA ansatz, which has the form of a linear 
combination of ``one particle plus one (conduction band) hole'' 
excitations over the ground state. To construct the intermediate 
states, $|int\rangle$, we need the HF (valence band) hole states. 
The latter are obtained by solving the Kohn-Luttinger Hamiltonian 
in the presence of the external confinement and the $N_e$-electron 
background. The RPA ansatz for the intermediate states has the
form of a linear combination of ``one electron (above the Fermi
level) plus one (valence band) hole'' excitations over the ground
state. As a direct result of the RPA calculations, we obtain 
matrix elements such as $\langle f|\hat e^\dagger_{\boldsymbol{\sigma}}
\hat e_{\boldsymbol{\lambda}}|i\rangle$, which are needed to compute the 
Raman amplitude.

\subsection{HF states for electrons}

We will model the qdot with a disk of thickness $L$. The disk axis
coincides with the $z$ axis. At $z=0$ and $z=L$ a hard wall potential
confines the electron motion. On the other hand, the in-plane
confining potential will be assumed to be parabolic
\cite{Hawrylak}, with a characteristic energy $\hbar\omega_0$. The
HF electron single-particle states (orbitals) are expanded in
terms of two-dimensional (2D) oscillator wave functions
\cite{PhysE2000}, $\varphi_s$, according to the following ansatz:

\begin{equation}
\phi_{\boldsymbol{\alpha}}(\vec r)=\sqrt{\frac{2}{L}}
 \sin (k_z^{\boldsymbol{\alpha}}\pi z/L)
 \sum_{s} C^{\boldsymbol{\alpha}}_{s} \varphi_s (x,y) \chi_s (\sigma),
\end{equation}

\noindent where $L$ is given in nanometers, $k_z$ is the sub-band
label, and $\chi_s$ are spin functions. 

The expansion coefficients, $C^{\boldsymbol{\alpha}}_s$, and the energy 
eigenvalues, $E_{\boldsymbol{\alpha}}$, are obtained from a set of 
equations similar to Eq. (6) of Ref. [\onlinecite{nuestro}], in which hole 
contributions shall be ignored:

\begin{widetext}
\begin{eqnarray}
\sum_t&&\left\{ E_{s,k_z^{\boldsymbol{\alpha}}}^{(0)}\delta_{st}+ 
 \beta\frac{e^2}{\kappa}\sum_{\boldsymbol{\gamma}\le \mu_F} 
 \sum_{u,v} \left[ \langle s,k_z^{\boldsymbol{\alpha}};
 u,k_z^{\boldsymbol{\gamma}}|\frac{1}{\sqrt{x^2+y^2}}|
 t,k_z^{\boldsymbol{\alpha}};v,k_z^{\boldsymbol{\gamma}}\rangle -
 \langle s,k_z^{\boldsymbol{\alpha}};u,k_z^{\boldsymbol{\gamma}}|
 \frac{1}{\sqrt{x^2+y^2}}|v,k_z^{\boldsymbol{\gamma}};
 t,k_z^{\boldsymbol{\alpha}}\rangle\right]
 C^{\boldsymbol{\gamma}}_u C^{\boldsymbol{\gamma}}_v \right\}
 C_{t}^{\boldsymbol{\alpha}}\nonumber\\
 &=&E_{\boldsymbol{\alpha}} C_{s}^{\boldsymbol{\alpha}},
\label{HFe}
\end{eqnarray}
\end{widetext}

\noindent
where $\boldsymbol{\gamma}$ runs over occupied states ($\mu_F$ is the 
electron Fermi level), and the 2D oscillator energies, in meV, are written 
as:

\begin{equation}
E^{(0)}_{s,k_z}=\frac{375.5~k_z^2}{(m_e/m_0) L^2}+\hbar\omega_0
 \{2 k_s+|l_s|+1\}.
\end{equation}

\noindent 
To be definite, we will use parameters appropriate for GaAs, 
i.e., the conduction band effective mass for electrons is $m_e/m_0=0.067$,
and the relative dielectric constant is $\kappa=12.5$.

Two-dimensional Coulomb matrix elements \cite{PhysE2000} will be
used instead of the truly 3D ones. Consequently, we will assume
that the matrix elements will be diagonal in the sub-band index,
$k_z$, and will multiply the matrix element by a strength
coefficient, $\beta$, in order to simulate the smearing effect of the $z$
direction \cite{MacDonald}. The coefficient will take values from 0.6 for $L=25$
nm, to 0.8 for $L=8$ nm.

The HF equations are solved iteratively. Twenty oscillator shells
are used in the calculations.

\subsection{HF states for holes}

To guarantee that both electrons and holes are confined in the
same spatial region, we will assume different confining
potentials, i.e., we will require:

\begin{equation}
m_e \omega_0=m^{hh}_{\parallel}\omega_0^{hh}=m^{lh}_{\parallel}
 \omega_0^{lh},
\end{equation}

\noindent where $m^{hh}_{\parallel}$ is the in-plane mass of the
$j=3/2$, $m_j=\pm 3/2$ (heavy) hole, and $m^{lh}_{\parallel}$ is
the mass of the $j=3/2$, $m_j=\pm 1/2$ (light) hole. The ansatz
for the HF hole orbitals is the following:

\begin{equation}
\phi_{\boldsymbol{\alpha}}^{(h)}(\vec r)=\sqrt{\frac{2}{L}}\sum_{s,k_z,m_j}
C^{\boldsymbol{\alpha}(h)}_{s,k_z,m_j}
 \sin\left(\frac{k_z\pi z}{L}\right) \varphi_s(x,y)~\chi_{m_j}.
\end{equation}

\noindent 
The expansion coefficients and energy eigenvalues are to be determined 
from the equations:

\begin{widetext}
\begin{eqnarray}
\sum_{t,k'_z,m'_j}&&\left\{\left(\cal{H}_{KL}\right)_{t,k'_z,m'_j}^
 {s,k_z,m_j}-\beta\frac{e^2}{\kappa}\sum_{\boldsymbol{\gamma}\le\mu_F}\,\sum_{u,v}
 \langle(s,k_z,m_j);(u,k_z^{\boldsymbol{\gamma}})|\frac{1}{\sqrt{x^2+y^2}}|
 (t,k'_z,m'_j);(v,k_z^{\boldsymbol{\gamma}})\rangle\; 
 C_{u}^{\boldsymbol{\gamma} (e)}
 C_{v}^{\boldsymbol{\gamma} (e)} \right\} 
 C_{t,k'_z,m'_j}^{\boldsymbol{\alpha} (h)}
 \nonumber\\
 &&= E_{\boldsymbol{\alpha}}^{(h)} C_{s,k_z,m_j}^{\boldsymbol{\alpha} (h)}.
\label{eq8}
\end{eqnarray}
\end{widetext}

The first term is the Kohn-Luttinger Hamiltonian, $\cal{H}_{KL}$,
whose matrix elements are given in Appendix
\ref{appendixA}. The second term is the electrostatic field of the
background $N_e$ electrons. Coulomb interactions are assumed to be
diagonal in $m_j$ indexes as well. Notice that, because
of the form of $\cal{H}_{KL}$, hole states are grouped into sets
with a common value of $f_h=-m_j+l_s$, where $l_s$ is the angular
momentum quantum number corresponding to the hole oscillator state
$\varphi_s$. 

HF energies and wavefunctions for electrons and
holes are used as input in the calculation of the many particle
functions $|f\rangle$ and $|int\rangle$.

\subsection{The final states}

The final states, $|f\rangle$, are excitations of the
$N_e$-electron system. In the RPA, they are obtained as linear 
combinations of ``one particle plus one hole'' excitations over
the initial state, $|i\rangle$:\cite{nuclear}

\begin{equation}
|f\rangle=\sum_{\boldsymbol{\lambda}\le\mu_F,\boldsymbol{\sigma}>\mu_F} 
 \left(X_{\boldsymbol{\sigma\lambda}}
 \hat e^\dagger_{\boldsymbol{\sigma}} \hat e_{\boldsymbol{\lambda}} - 
 Y_{\boldsymbol{\lambda\sigma}}
 \hat e^\dagger_{\boldsymbol{\lambda}} \hat e_{\boldsymbol{\sigma}}\right) |i\rangle,
 \label{eq9}
\end{equation}

\noindent where the index $\boldsymbol{\lambda}$ runs over occupied HF states,
and $\boldsymbol{\sigma}$ runs over unoccupied states. Detailed equations for
the coefficients $X$, $Y$ and the energy eigenvalues can be straightforwardly
obtained from the formulas (12-15) of Ref. [\onlinecite{nuestro}]:

\begin{eqnarray}
\sum_{\boldsymbol{\tau,\mu}}\left\{ A_{\boldsymbol{\sigma\lambda,\tau\mu}} 
 X_{\boldsymbol{\tau\mu}}+B_{\boldsymbol{\sigma\lambda,\mu\tau}} 
 Y_{\boldsymbol{\mu\tau}}\right\}&=& \hbar\Omega_f X_{\boldsymbol{\sigma\lambda}},
 \nonumber\\ \sum_{\boldsymbol{\tau,\mu}}\left\{ 
 B_{\boldsymbol{\lambda\sigma,\tau\mu}} X_{\boldsymbol{\tau\mu}}
 +A_{\boldsymbol{\lambda\sigma,\mu\tau}} Y_{\boldsymbol{\mu\tau}}\right\}&=&
 -\hbar\Omega_f Y_{\boldsymbol{\lambda\sigma}},
\label{RPAeq}
\end{eqnarray}

\noindent
in which $\hbar\Omega_f$ is the excitation energy,  $\boldsymbol{\tau}$ and 
$\boldsymbol{\mu}$ are indexes similar to $\boldsymbol{\sigma}$ and 
$\boldsymbol{\lambda}$, respectively, and the $A$ 
and $B$ matrices are given by:

\begin{widetext}
\begin{eqnarray}
A_{\boldsymbol{\sigma\lambda,\tau\mu}} &=& (E^{(e)}_{\boldsymbol{\sigma}}-
 E^{(e)}_{\boldsymbol{\lambda}}) \delta_{\boldsymbol{\sigma\tau}} 
 \delta_{\boldsymbol{\lambda\mu}}
 + \frac{\beta e^2}{\kappa} \left( \langle \boldsymbol{\sigma,\mu} |
 \frac{1}{\sqrt{x^2+y^2}}|\boldsymbol{\lambda,\tau}\rangle -
 \langle \boldsymbol{\sigma,\mu} |\frac{1}{\sqrt{x^2+y^2}}|
 \boldsymbol{\tau,\lambda}\rangle \right),\nonumber\\
B_{\boldsymbol{\sigma\lambda,\mu\tau}} &=& \frac{\beta e^2}{\kappa}
 \left( \langle \boldsymbol{\sigma,\tau}
 |\frac{1}{\sqrt{x^2+y^2}}|\boldsymbol{\lambda,\mu}\rangle -
 \langle \boldsymbol{\sigma,\tau} | \frac{1}{\sqrt{x^2+y^2}}|
 \boldsymbol{\mu,\lambda}\rangle \right). 
\end{eqnarray}
\end{widetext}

Notice that $X_{\boldsymbol{\sigma\lambda}}$ has a straightforward 
interpretation in terms of a transition amplitude:

\begin{equation}
X_{\boldsymbol{\sigma\lambda}}=\langle i|\hat e_{\boldsymbol{\lambda}}^\dagger 
 \hat e_{\boldsymbol{\sigma}}|f\rangle,
\end{equation}

\noindent and similarly for the $Y_{\boldsymbol{\lambda\sigma}}$.

We shall stress that final states are characterized by the
quantum numbers $\Delta l$ and $\Delta S_z$ representing the
variation, with respect to the ground state, of the total angular
momentum projection and the total spin projection, respectively.
Conventionally, we will call $\Delta l=0$ states as monopole
excitations, $\Delta l=\pm 1$ states as dipole excitations, 
$\Delta l=\pm 2$ as quadrupole excitations, etc.

The calculation of strength functions defined in Eq. (\ref{eq3})
follows also very simply from the results of Ref.
[\onlinecite{nuestro}]. One expands the exponential

\begin{eqnarray}
e^{i (\vec q_i-\vec q_f)\cdot \vec r}&=& e^{i (q_{zi}-q_{zf}) z}
 \;e^{i(\vec q_{\parallel i}-\vec q_{\parallel
 f})\cdot\vec\rho}\nonumber\\
 &=&e^{i (q_{zi}-q_{zf}) z}\;\left(1+i
 (\vec q_{\parallel i}-\vec q_{\parallel
 f})\cdot\vec\rho+\dots\right),\nonumber\\
\end{eqnarray}

\noindent and makes use of the definition of multipole operators,
$d^l_{\alpha\gamma}$, given in Ref. [\onlinecite{nuestro}]:
 
\begin{eqnarray}
d^l_{\alpha\gamma}&=&\langle \alpha|\rho^{|l|} e^{i l \theta}
 |\gamma\rangle;~~l\ne 0,\nonumber \\
&=&\langle \alpha|\rho^2|\gamma\rangle;~~l=0,
\end{eqnarray}

\noindent
whose detailed expressions can be found in Appendix B of that
reference. In the later formulas, $\alpha$ denotes the orbital
part (no spin function included) of the HF electronic state
$\boldsymbol{\alpha}$. The spin projection quantum number is
explicitly indicated in Eq. (\ref{eq3}).
With respect to the part depending on $z$, one uses that

\begin{equation}
\langle k_z|e^{i q z}|k'_z\rangle=\frac{4 i k_z k'_z q \left(
 -1+e^{i\pi q}\cos (k_z\pi) \cos (k'_z\pi)\right)}{\pi \left[
 (k_z-k'_z)^2-q^2\right]\left[(k_z+k'_z)^2-q^2\right]}.
 \label{eq11}
\end{equation}

The strength functions, or more precisely the multipole operators,
allow a further classification of the $|f\rangle$ states into
collective and single-particle excitations\cite{nuestro}. A
charge monopolar collective state $|f\rangle$, for example, 
gives a significantly nonzero value for the matrix element:

\begin{eqnarray}
D^0_{fi}=\left\langle f\left| \sum_{\alpha,\alpha'}
 d^0_{\alpha,\alpha'}[ \hat e^\dagger_{\alpha\uparrow}\hat e_{\alpha'\uparrow}+
 \hat e^\dagger_{\alpha\downarrow}\hat e_{\alpha'\downarrow} ]
 \right|i\right\rangle,
\end{eqnarray}

\noindent whereas for a single-particle excitation, the matrix
element practically equals zero. By ``significantly nonzero value''
we mean that $|D^0_{fi}|^2$ is greater than 5\% of the 
energy-weighted sum rule for the monopole operator
\cite{nuclear,nuestro}:

\begin{equation}
\sum_f \hbar\Omega_f |D^0_{fi}|^2=\frac{2\hbar^2}{m_e}
 \sum_{\boldsymbol{\lambda}\le\mu_F} \langle \boldsymbol{\lambda}|\rho^2|
 \boldsymbol{\lambda} \rangle.
\end{equation}

Similar criteria can be formulated for charge multipolar states
\cite{nuestro}, or for spin excited states (involving or not
spin reversal with respect to the ground state). The latter are
related to the last three terms of Eq. (\ref{eq3}).

\subsection{The intermediate states}

The intermediate states with $N_e+1$ electrons and one
hole, can be obtained from the so-called particle-particle
Tamm-Dankoff approximation (pp-TDA), which is an uncorrelated pp-RPA
function, i.e., no particles below the Fermi level are created 
\cite{nuclear}:

\begin{equation}
|int\rangle=\sum_{\boldsymbol{\sigma} > \mu_F,\boldsymbol{\tau}} 
 V_{\boldsymbol{\sigma\tau}} \hat e^\dagger_{\boldsymbol{\sigma}} 
 \hat h^\dagger_{\boldsymbol{\tau}} \;|i\rangle,
 \label{eq12}
\end{equation}

\noindent where $\boldsymbol{\sigma}$ is a HF electron state above the
Fermi level, and $\boldsymbol{\tau}$ a HF hole state.
The equations for the expansion coefficients, $V$, and
the energy eigenvalues are explicitly written in Ref.
[\onlinecite{nuestro}]:
 
\begin{eqnarray}
(\hbar\Omega_{int}&-&E^{(e)}_{\boldsymbol{\sigma}}-E^{(h)}_{\boldsymbol{\tau}}) 
 V_{\boldsymbol{\sigma\tau}} = \nonumber\\
 &-&\frac{\beta e^2}{\kappa}\sum_{\boldsymbol{\sigma',\tau'}} \langle 
 \boldsymbol{\sigma,\tau}|\frac{1}{\sqrt{x^2+y^2}}|\boldsymbol{\sigma',\tau'}\rangle
 V_{\boldsymbol{\sigma'\tau'}}.
\label{pp-RPA}
\end{eqnarray}

\noindent
The quantity $\hbar\Omega_{int}$ gives the excitation energy, measured with
respect to the ground state of the $N_e$-electron system, and the coefficients
$V_{\boldsymbol{\sigma\tau}}$ can be interpreted as the transition amplitudes:

\begin{equation}
V_{\boldsymbol{\sigma\tau}}=\langle i|\hat h_{\boldsymbol{\tau}}
 \hat e_{\boldsymbol{\sigma}}|int\rangle.
\end{equation}
 
The intermediate states are characterized by the quantum numbers

\begin{equation}
{\cal F}=l_e+f_h, \quad S_z,
\end{equation}

\noindent where $l_e$ and $S_z$ are the angular momentum and spin
projection of the added electron.

\subsection{The geometry of the Raman experiment}

In the present paper, we restrict ourselves to the so-called
backscattering geometry, which is often used in experiments.
The incident laser beam is deflected inside the dot because 
of Snell's law. Thus, the actual angle of incidence 
(with respect to the $z$ axis) is:

\begin{equation}
\phi'_i=\arcsin \left( \frac{\sin \phi_i}{\eta}\right),
\end{equation}

\noindent where $\eta\approx 3.5$ is the qdot refractive index.
If $q_i$ denotes the wave vector in vacuum, then inside the dot
we have:

\begin{eqnarray}
q'_{\parallel i}&=&q_{\parallel i}= q_i \sin \phi_i,\\
q'_{zi}&=&\eta q_i\cos \phi'_i,
\end{eqnarray}

\noindent
and for the scattered light:

\begin{eqnarray}
q'_{\parallel f}&=& q_f \sin \phi_f,\\
q'_{zf}&=&-\eta q_f\cos \phi'_f,
\end{eqnarray}

\noindent
with $\phi_i=\phi_f$.

We distinguish between the polarized geometry, in which
$\vec\varepsilon_i$ and $\vec\varepsilon_f$ are parallel (along
the $y$ axis in our calculations), and the depolarized geometry,
in which $\vec\varepsilon_f$ (in the $xz$ plane) is orthogonal to
$\vec\varepsilon_i$. A detailed definition of angles for more
general geometries can be found in [\onlinecite{nuestro}].

\subsection{The matrix elements of $\hat H_{e-r}$}

$\hat H_{e-r}^-$ is the part of the electron-radiation Hamiltonian
responsible for the annihilation of an incident photon and
creation of an e-h pair. Its matrix elements are written as
\cite{nuestro}:

\begin{equation}
\langle int|\hat H^-_{e-r}|i\rangle\sim\sum_{\boldsymbol{\sigma}>\mu_F,
 \boldsymbol{\tau}} (band-orbital)_{\boldsymbol{\sigma\tau}}^{(i)}\;
 V_{\boldsymbol{\sigma\tau}}^*,
\label{eq15}
\end{equation}

\noindent where $V_{\boldsymbol{\sigma\tau}}$ are the coefficients entering the
pp-TDA Eq. (\ref{eq12}). Because of valence-band
mixing, the orbital (envelope) and band wave functions of the hole
get mixed. The band-orbital factor in Eq. (\ref{eq15}) is defined
as:

\begin{widetext}
\begin{eqnarray}
(band-orbital)_{\boldsymbol{\sigma\tau}}^{(i)}=\sum_{s,k_z}\;\sum_{t,k'_z,m_j}
 C_{s,k_z}^{\boldsymbol{\sigma} (e)*} C_{t,k'_z,m_j}^{\boldsymbol{\tau} (h)*}
 \left(\vec\varepsilon_i\cdot\vec p_{\boldsymbol{\sigma}_s,m_j} \right)
 \langle k_z|e^{i q_{zi} z}|k'_z\rangle
 \;\int e^{i\vec q_{\parallel i}\cdot\vec\rho}
 \varphi_{se}^*(\vec\rho) \varphi_{th}^*(\vec\rho)\;{\rm d}^2\rho.
 \label{eq16}
\end{eqnarray}
\end{widetext}

The band factor, $\vec\varepsilon_i\cdot\vec p_{\boldsymbol{\sigma}_s,m_j}$, is
computed according to Table \ref{tab1}. The factor $\langle
k_z|e^{i q_{zi} z}|k'_z\rangle$ is computed with the help of Eq.
(\ref{eq11}). Finally, the computation of the integral (the
orbital factor) is made along the lines sketched in Ref.
[\onlinecite{nuestro}].

On the other hand, $\hat H_{e-r}^+$ is that part of the
electron-radiation Hamiltonian responsible for the creation of a
photon and annihilation of an e-h pair. Its matrix elements are
given as \cite{nuestro}:

\begin{equation}
\langle f|\hat H^+_{e-r}|int\rangle\sim\sum_{\boldsymbol{\sigma}>\mu_F,
 \boldsymbol{\lambda}\le\mu_F} \sum_{\boldsymbol{\tau}}
 (band-orbital)_{\boldsymbol{\lambda\tau}}^{(f)} V_{\boldsymbol{\sigma\tau}}
 X_{\boldsymbol{\sigma\lambda}}^*,
\end{equation}

\noindent where $X_{\boldsymbol{\sigma\lambda}}$ is one of the coefficients
entering the RPA expansion Eq. (\ref{eq9}). The band-orbital
factor is obtained from Eq. (\ref{eq16}) by replacing $i$ by $f$
and taking the complex conjugate of the expression.

\subsection{Phenomenological $\Gamma_f$ and $\Gamma_{int}$}

The main decay mechanism of electronic excited levels in a qdot at
very low temperatures is the emission of longitudinal optical (LO)
phonons \cite{Shah}. We will ignore surface effects in a qdot, and
assume a threshold excitation energy appropriate for GaAs,
$\hbar\omega_{LO}\approx 30$meV, for the emission of LO phonons.

Only final states with excitation energies lower than
$\hbar\omega_{LO}$ will be considered in order to exclude Raman
peaks related to phonon excitations. It means that final states
will have small widths, for which we will take a constant value,
$\Gamma_f$, in the interval between 0.1 and 0.5 meV.

In the same way, for intermediate states with excitation energy
lower than $\hbar\omega_{LO}$ we will take $\Gamma_{int}=$0.5 meV.
For higher excitation energies, the LO phonon decay mechanism
becomes active and the widths suddenly increase. In this case, we
will take $\Gamma_{int}=$10 meV, except for a set of particular
states, which can be interpreted as ``excitons'' or ``excitons +
plasmons'', whose meaning will become clear below in the
discussion of Raman scattering well above band gap. In this
latter situation, we will take $\Gamma_{int}=$2 meV.

We stress that the role of $\Gamma_{f}$ and $\Gamma_{int}$ as
functions of the excitation energy in the Raman spectra has not
been pointed out before. In our view, the qualitative change of
the resonant Raman spectrum when the incident laser energy,
$h\nu_i$, is raised 30 or more meV above the band gap is related
to the sudden increase of $\Gamma_{int}$.

\section{Calculated Raman spectra}
\label{sec3}

In the following, we report results for a 42-electron dot. The
disk thickness and the harmonic confinement take values $L=25$ nm,
and $\hbar\omega_0=12$ meV,\cite{nota2} respectively. The chosen
$\hbar\omega_0$ corresponds to a qdot in the strong confinement
regime, and the number of electrons to a closed-shell quantum dot.

\begin{figure*}[h]
\begin{center}
\includegraphics[width=.7\linewidth,angle=-90]{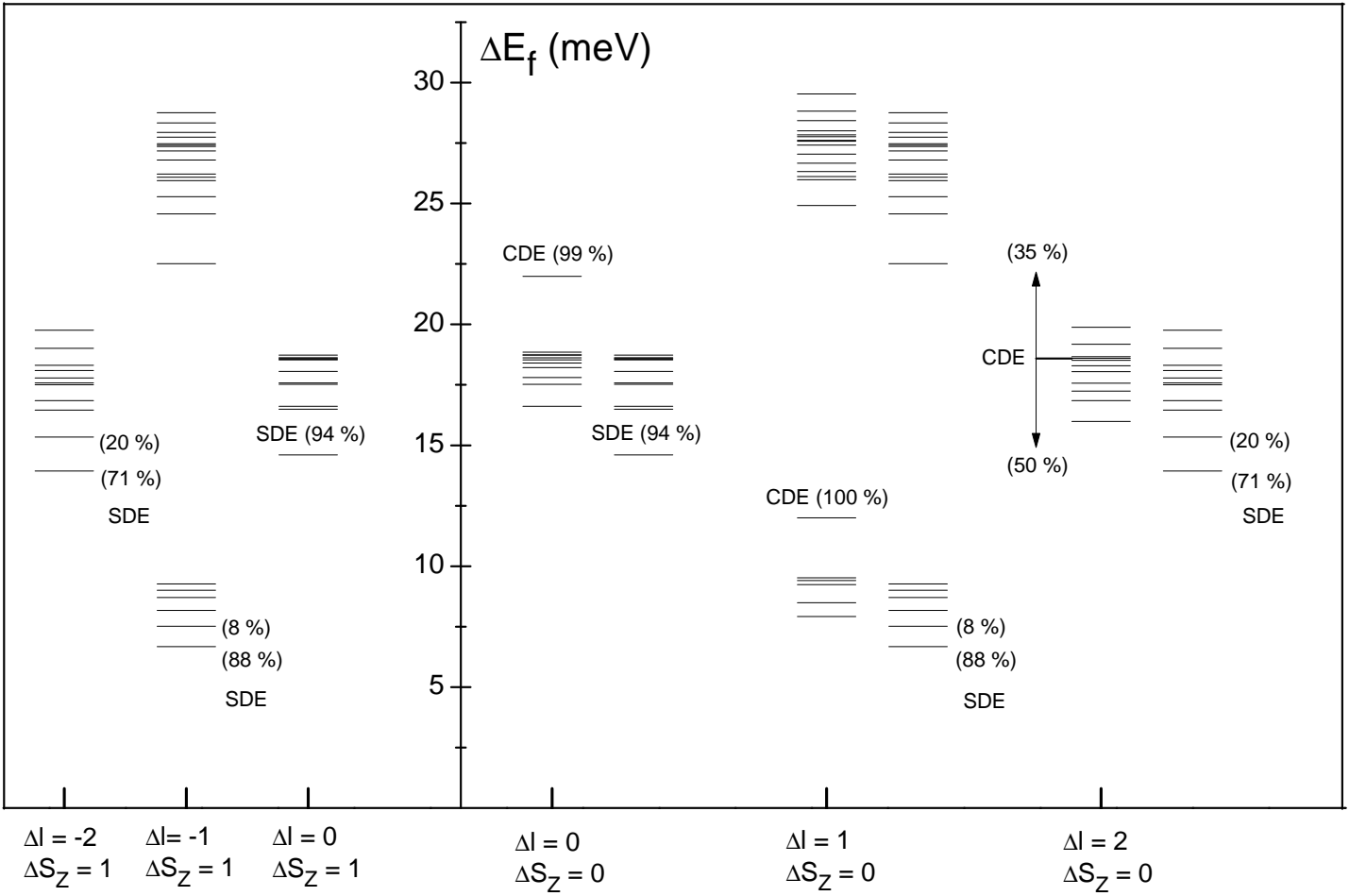}
\caption{\label{fig2} Spectrum of final states in the model
 qdot.}
\end{center}
\end{figure*}

We show in Fig. \ref{fig2} the electronic excitations of the dot,
i.e., the spectrum of final states, $|f\rangle$. The reference
energy is the energy of the ground state, $|i\rangle$. The
excitation energy is precisely what is measured as the Raman shift
in the experiments.

To the left of the $y$ axis, states with $\Delta S_z=1$ (with
respect to $|i\rangle$) are represented, while to the right of the
$y$ axis, states with no spin flips are shown. In the figure, we
identify the collective excitations, labelled CDE and SDE, and
give explicitly the corresponding fraction of the energy-weighted
sum rule \cite{nuestro}. In the $\Delta l=0$, $\Delta S_z=0$ case, for example,
the CDE state concentrates the strength of the charge monopolar
transition (from $|i\rangle$ to $|f\rangle$), and the SDE state
concentrates the strength of the spin monopolar transition (with
no spin flip). The rest of the states shown correspond to
single-particle excitations (SPE). Notice that, in general,
collective excitations are isolated from SPEs.

For the intermediate states, we take a nominal band gap,
$E_{gap}=1560$ meV. This gap is renormalized by Coulomb
interactions. We define the renormalized $E'_{gap}$ as the energy
of the lowest intermediate state. Note that this convention may
not coincide with the experimental definition of the effective gap
in terms of the position of the exciton line.

\subsection{Raman spectra below the effective band gap}

Measurements of electronic Raman spectra when the laser excitation
energy, $h \nu_i$, is below $E'_{gap}$ have not, to the best of
our knowledge, been reported for qdots. In the present section,
however, we show that such measurements could provide information
for both collective excitations and SPEs in qdots. Raman
intensities for both kinds of excitations show comparable
magnitudes.

Note that we use only the resonant contribution to $A_{fi}$, Eq.
(\ref{eq1}), in spite of the fact that the present situation does
not correspond, strictly speaking, to a resonant process.
\cite{nota}

We have the possibility of computing the spectrum for each
multipolarity of final states. Results will be presented in this
way, although in an experiment all the multipolarities can be
observed in the same spectrum.

\begin{figure*}[ht]
\begin{center}
\includegraphics[width=.8\linewidth,angle=-90]{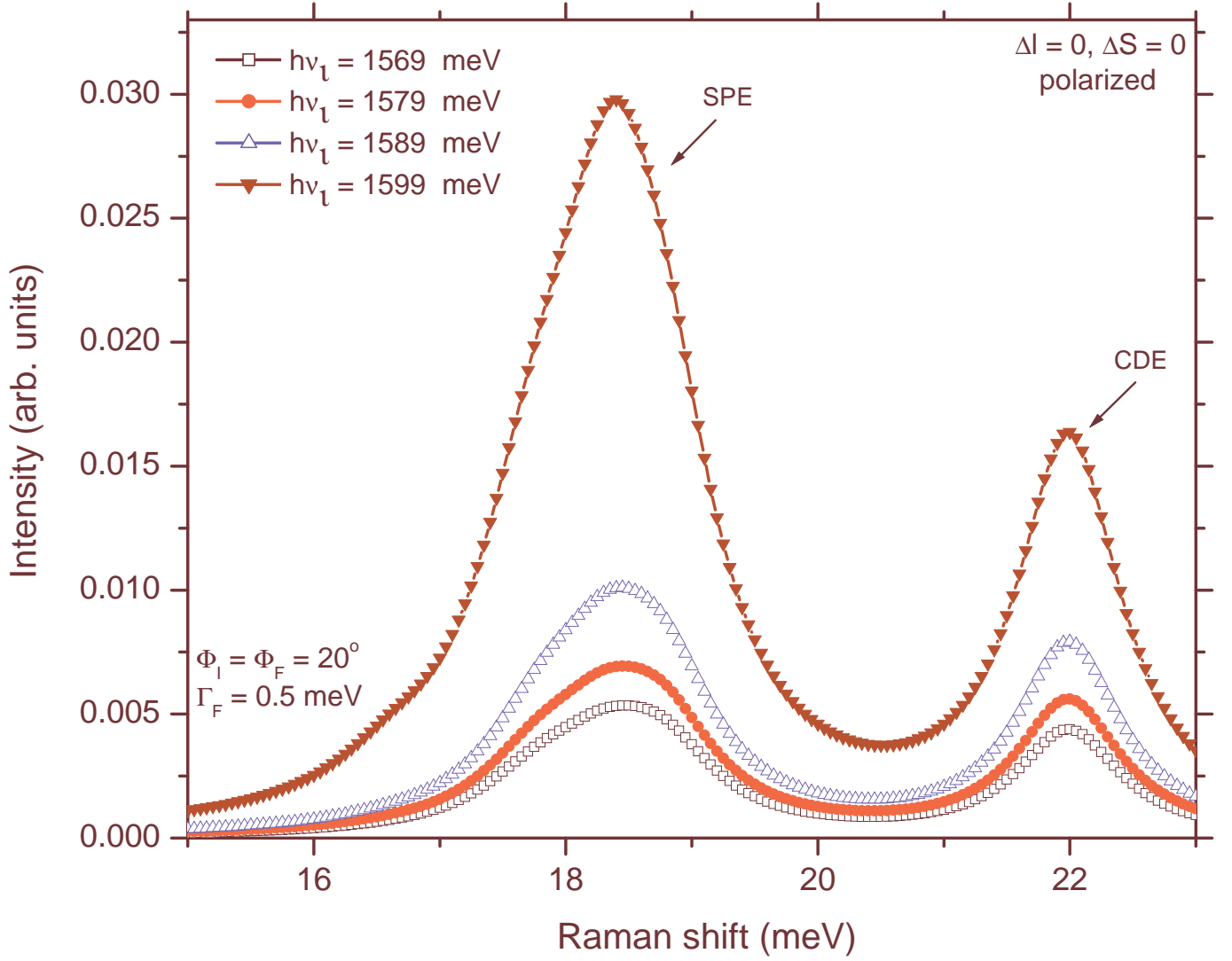}
\caption{\label{fig3} (Color online) Polarized monopolar Raman spectrum 
 below  band gap ($\Gamma_f= 0.5$ meV).}
\end{center}
\end{figure*}

We show in Fig. \ref{fig3} the polarized Raman spectrum for
monopole final state excitations, computed with $\Gamma_f=0.5$
meV. $E'_{gap}$ in this situation is 1599.2 meV. The incident (and
backscattered) angle is equal to $20^\circ$. $h\nu_i$ is swept in
a 30 meV interval below $E'_{gap}$. Notice the monotonic increase
of intensities as $h\nu_i$ rises. One peak corresponding to the
CDE, and a second one related to the SPEs are observed. In the
latter case, there is a group of energy levels contributing to the
peak in the figure. We may think of this set of levels as a
Coulomb-renormalized oscillator shell. As we are dealing with
monopole excitations, the average position of the group should
correspond to $2\hbar\omega'_0$, i.e., the renormalized oscillator
energy is $\hbar\omega'_0\approx9$ meV.

\begin{figure*}[ht]
\begin{center}
\includegraphics[width=.9\linewidth,angle=0]{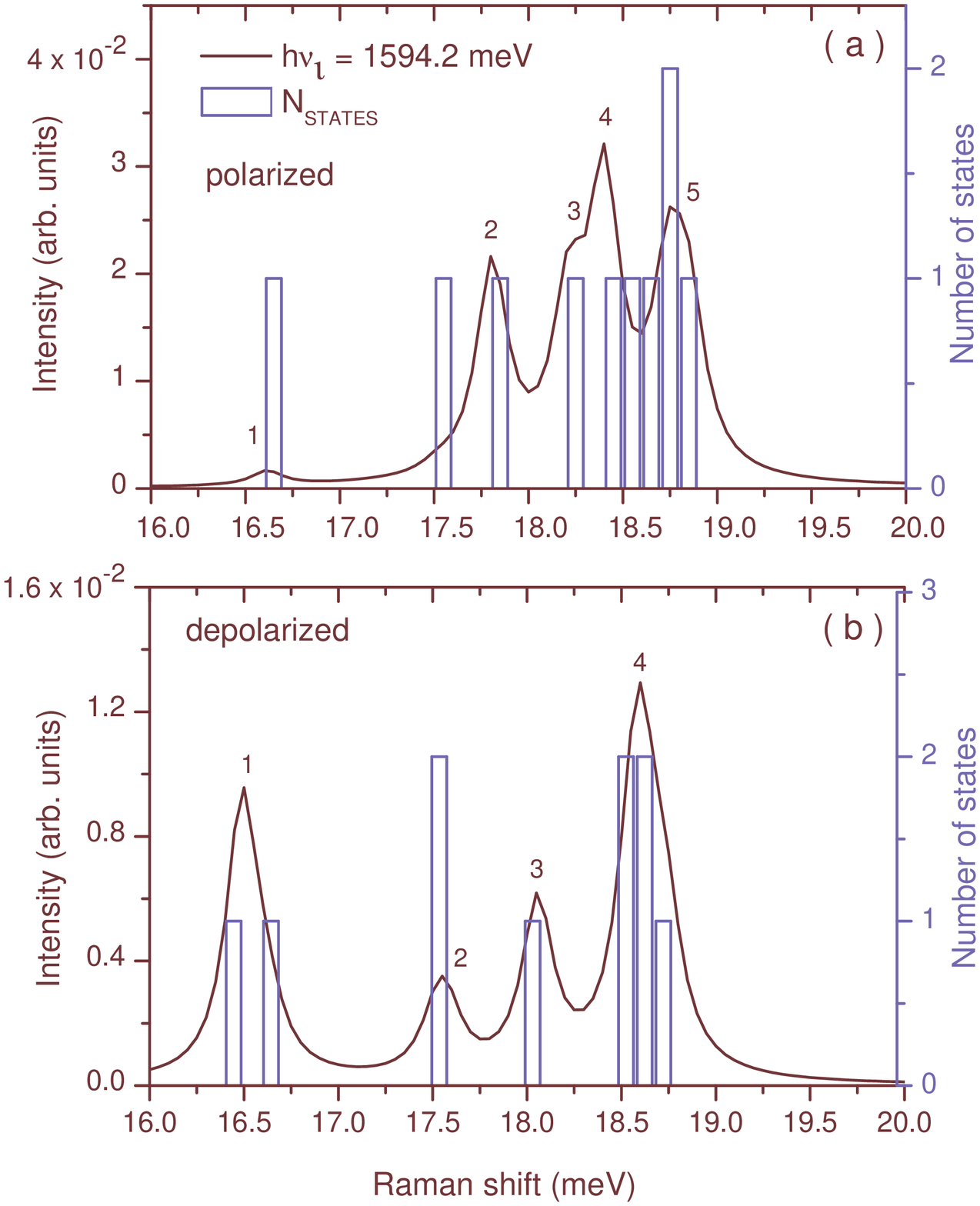}
\caption{(Color online) Polarized and depolarized monopolar Raman spectra
($\Gamma_f=0.1$ meV) and comparison with the density of energy
levels. \label{fig4}}
\end{center}
\end{figure*}

The fine structure of the SPE peak is shown in Fig. \ref{fig4}a
along with the density of final-state energy levels. In this case,
the monopolar polarized Raman spectrum was calculated with
$\Gamma_f=0.1$ meV. The depolarized spectrum is shown in Fig.
\ref{fig4}b (In fact, only the energy interval corresponding to
the SPE peak is shown. The SDE peak is outside this interval).
Histograms with a step of 0.1 meV are used to represent the level
density. Although Eq. (\ref{eq3}) refers to collective
excitations, we have used its implications to correlate the
polarized Raman spectrum with the level density of charge
monopolar SPEs, and the depolarized spectrum with the density of
monopolar spin excitations. The Raman spectra reproduce quite
accurately the details of the level density in both cases. A
general remark concerning Fig. \ref{fig4} is that the intensity of
the depolarized peaks is about three times lower than the
intensity of the polarized ones.

\begin{figure*}[ht]
\begin{center}
\includegraphics[width=.75\linewidth,angle=-90]{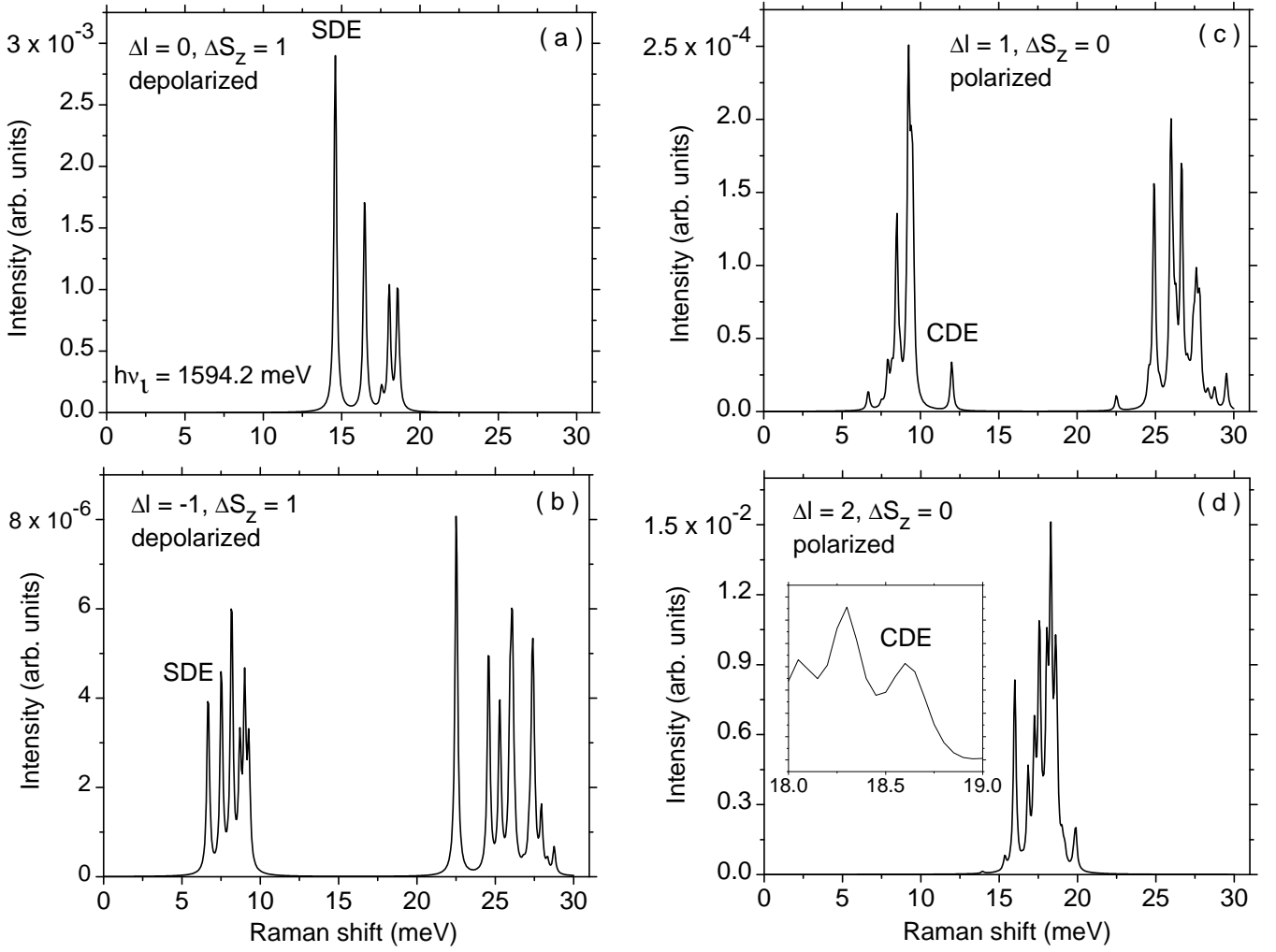}
\caption{Raman spectra in different channels ($\Gamma_f= 0.1$
 meV). The incident laser energy is $1594.2$ meV. \label{fig5}}
\end{center}
\end{figure*}

Depolarized spectra for spin-flipped monopolar and dipolar final
states are shown in Figs. \ref{fig5}a and b. Polarized spectra for
dipolar and quadrupolar states are shown in Figs. \ref{fig5}c and
d. The shell structure \cite{Lockwood2,Heitmann}is clearly seen in
these figures. Dipolar and spin-flipped final states are strongly
depressed in the Raman spectra. Quadrupole final states, on the
contrary, show magnitudes comparable to monopolar peaks.

The fact that even multipoles are favored in the Raman spectra is
understood in terms of the even parity of final states in a
two-photon process. On the other hand, spin-flipped final states
are reached only as a consequence of valence-band mixing. We
(virtually) create an electron and a hole, whose dominant
component is $m_j$, and annihilate the same hole and an electron
with opposite spin. The amplitude of the latter process is
proportional to the minority component of the hole wave function,
$\chi_{m'_j}$. It means that the amplitude squared $|A_{fi}|^2$
will be proportional to $|\chi_{m'_j}|^2$.

These calculations show that experimental measurements of the
electronic Raman spectra with below band gap excitation can
provide valuable information on the collective states and SPEs of
qdots. Furthermore, below band gap excitation can overcome the
problem of overlap with the intense photoluminescence observed
under resonant excitation. The peak maxima exhibit a continuous
but not very marked increase in intensity with excitation
approaching the band gap (see Fig. \ref{fig3}), indicating that
excitation around 30 meV below the gap is sufficient. The other
notable feature of these calculations, apart from the marked
differences predicted in Raman intensities of the polarized and
depolarized multipolar components, is the fine structure of the
SPE Raman peak (see Fig. \ref{fig4}). It would be interesting to
probe all these aspects experimentally.

\subsection{The extreme resonance region}

In the present section, we consider Raman scattering when $h\nu_i$
moves in a 30 meV window above $E'_{gap}$. We will call this
interval the ``extreme resonance'' window.

\begin{figure*}[ht]
\begin{center}
\includegraphics[width=.8\linewidth,angle=-90]{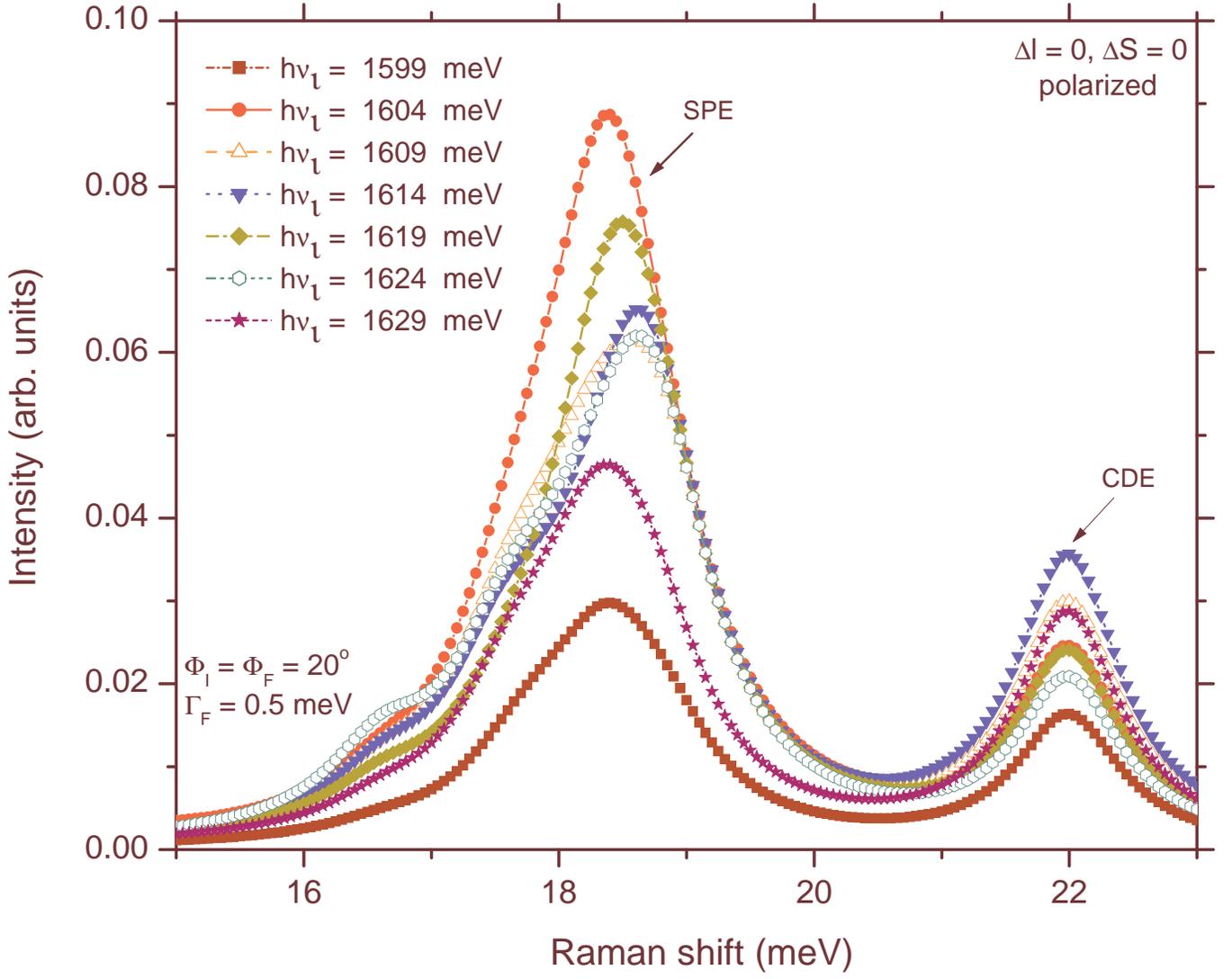}
\caption{(Color online) Monopolar polarized Raman spectra ($\Gamma_f= 0.5$ 
 meV) in the extreme resonance region: $E'_{gap}< h\nu_i < E'_{gap}+30$
 meV. \label{fig6}}
\end{center}
\end{figure*}

Fig. \ref{fig6} shows a polarized Raman spectrum corresponding to
charge monopolar final states. As in Fig. \ref{fig3}, we used
$\Gamma_f=0.5$ meV. Two characteristics of Fig. \ref{fig6} make it
very different from Fig. \ref{fig3}: (i) Peak intensities are not
monotonous with respect to variations of $h\nu_i$, and (ii) The
position of the maximum in the SPE peak moves slightly with
$h\nu_i$. Both properties are related to resonances in the
intermediate states.

\begin{figure*}[ht]
\begin{center}
\includegraphics[width=.9\linewidth,angle=0]{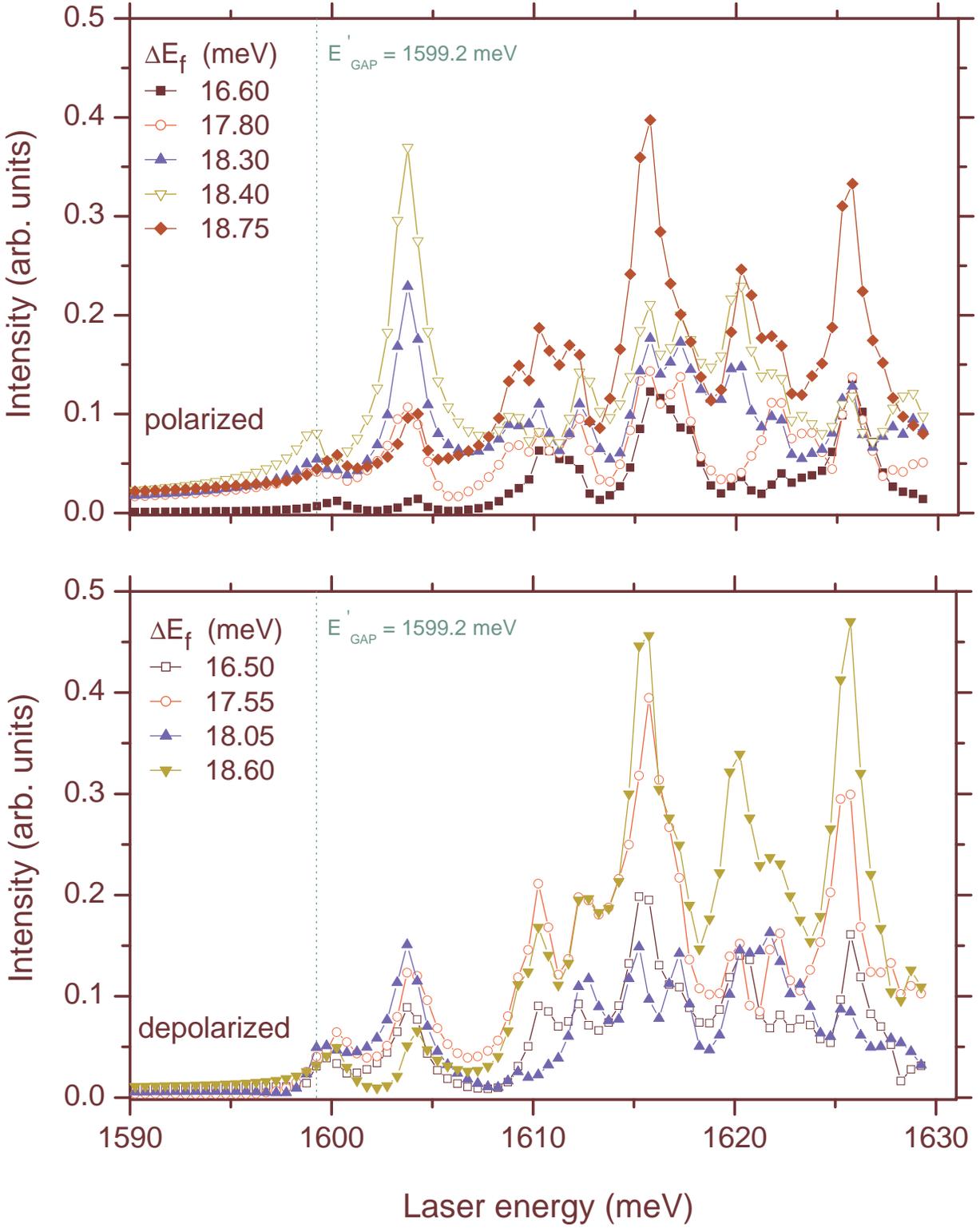}
\caption{(Color online) The dependence on $h\nu_i$ of the intensity of SPE 
 peaks identified in Fig. \ref{fig4}. \label{fig7}}
\end{center}
\end{figure*}

Resonances in the intermediate states can be better visualized if
we follow the Raman intensities of the peaks shown in Fig.
\ref{fig4}. For this purpose, we computed monopolar Raman spectra
with $\Gamma_f=0.1$ meV and varying $h\nu_i$ with a 0.5 meV step.
The results are drawn in Fig. \ref{fig7}. The monotonous increase
of peaks in the $h\nu_i < E'_{gap}$ region is apparent in the
figure. On the other hand, for $h\nu_i > E'_{gap}$ the intensity
variation with laser excitation energy is much more complicated.
The intensities of the individual SPE components rise and fall
markedly with laser energy, as has been observed experimentally. 
This variation is attributed to individual
resonances occurring within intermediate states lying close to the
band gap as the incident light energy sweeps through them.
The effect is particularly noticeable for $h\nu_i\approx 1616$ and 1626
meV. In the associated intermediate states, the added e-h pair has
zero total angular momentum projection, and the hole is basically
a heavy hole. Notice that the same intermediate states are
responsible for the strong enhancement of Raman intensities in
both the polarized and depolarized geometries.

\begin{figure*}[ht]
\begin{center}
\includegraphics[width=.9\linewidth,angle=0]{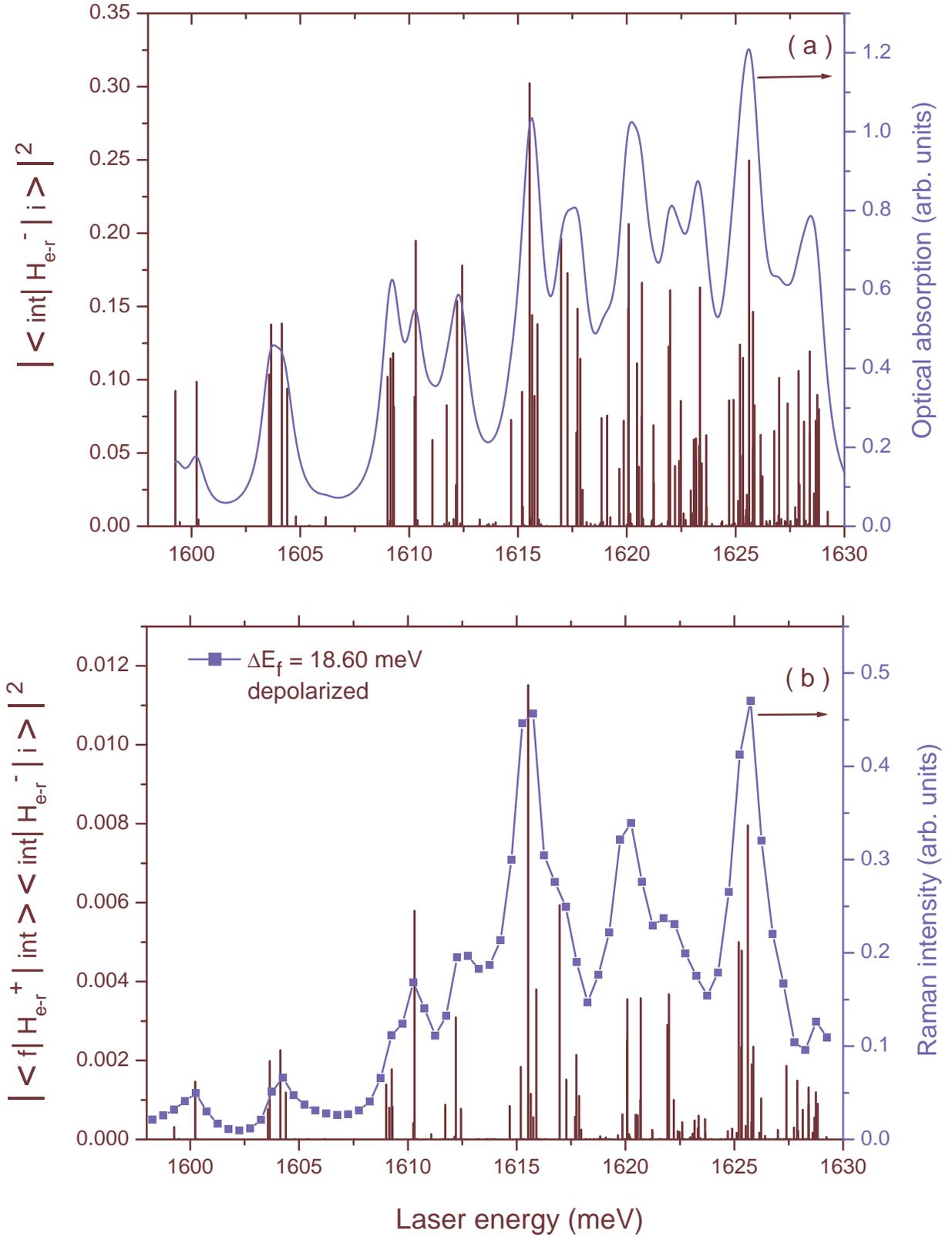}
\caption{(Color online) (a) Absorption in the model qdot. (b) Intensity of 
the spin monopolar Raman peak with $\Delta E_f=18.6$ meV as a function of 
$h\nu_i$. 
\label{fig8}}
\end{center}
\end{figure*}

A spin monopolar SPE Raman peak is followed as a function of $h\nu_i$ 
in Fig. \ref{fig8}b. For comparison, we have also given the
products $|\langle f |\hat H^+_{e-r}|int\rangle\langle
int|\hat H^-_{e-r}|i\rangle|^2$ for each intermediate state, and the
absorption strengths $|\langle int |\hat H^-_{e-r}|i\rangle|^2$ (the
upper panel). The optical absorption coefficient is defined according to

\begin{equation}
\alpha(E)=\sum_{int} |\langle int|\hat H^-_{e-r}|i\rangle|^2 \frac{\Gamma_{int}/\pi}
{(E-E_{int})^2+\Gamma_{int}^2}.
\end{equation}

\noindent
Fig. \ref{fig8} shows that, in the present situation, 
peaks in the optical absorption coincide with peaks in the Raman intensities, which
leads to the conclusion that the latter are related to incoming (absorption) 
resonances. 
Note that from Fig. \ref{fig8}b it follows that interference effects among 
intermediate states are weak in the Raman scattering under extreme resonance\cite{JPCM}.
Raman spectra in the extreme resonance region look similar to the
spectra shown in Fig. \ref{fig5}, but with SPE peaks much higher than collective ones and  
selectively enhanced for particular values of $h\nu_i$. A comparison with
the density of energy levels would lead to results very similar to those of Fig.
\ref{fig4}.

\subsection{Raman spectra 40 meV above band gap}

\begin{figure*}[ht]
\begin{center}
\includegraphics[width=.8\linewidth,angle=-90]{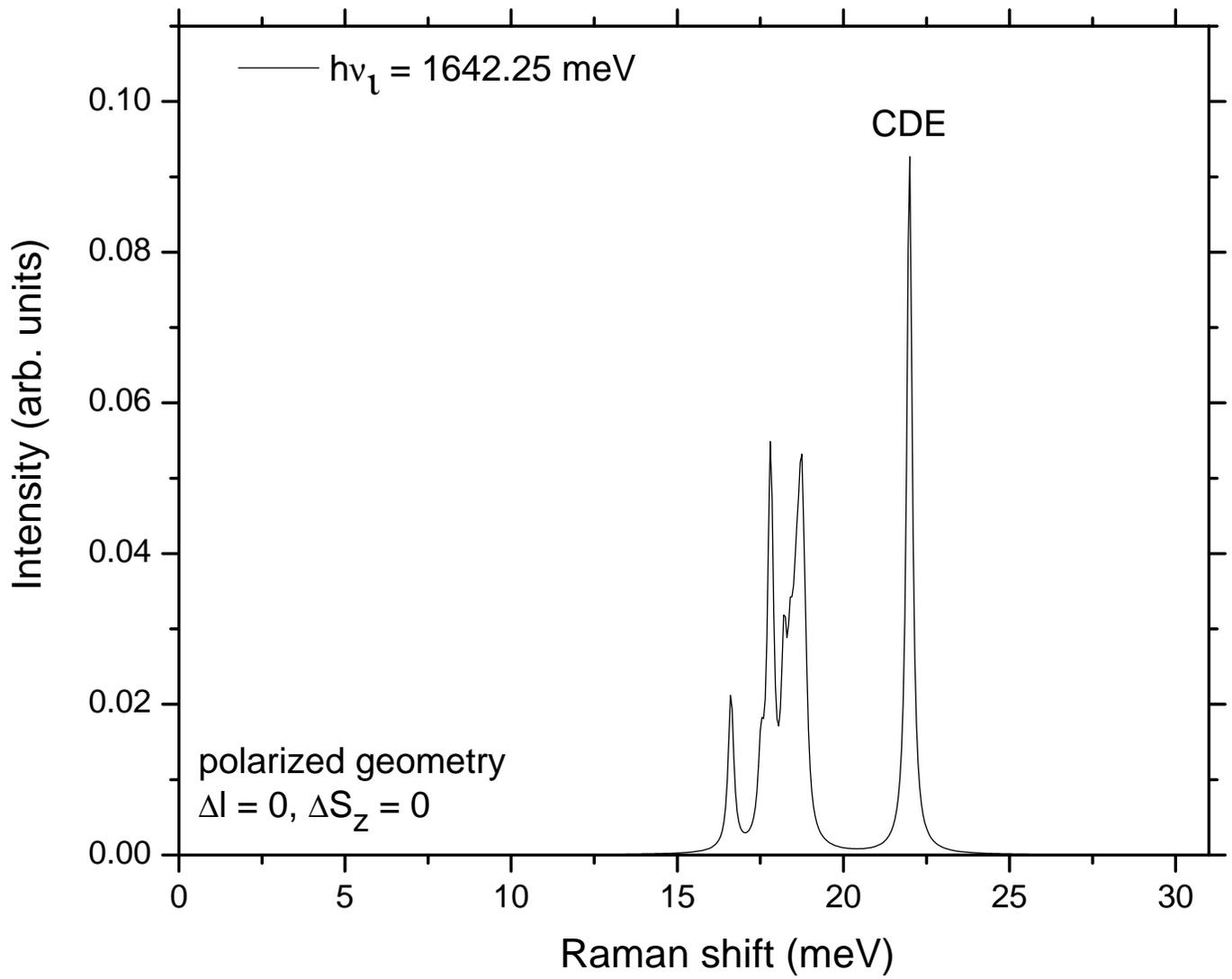}
\caption{Raman spectrum in the polarized geometry ($\Gamma_f=0.1$ meV) for well
above band gap excitation. \label{fig9}}
\end{center}
\end{figure*}

As mentioned in Section \ref{sec2}G, we assume that $\Gamma_{int}$ experiences a sudden 
increase when $E_{int} > E'_{gap}+\hbar\omega_{LO}$. As a result, the contribution of
these states to $A_{fi}$, Eq. (\ref{eq1}), loses its resonant character even when $h\nu_i$
sweeps this energy range. It means that peak intensities become smooth functions of
$h\nu_i$, as for below band gap excitation. Both collective and SPE Raman peaks decrease
in intensity for $h\nu_i > E'_{gap}+\hbar\omega_{LO}$ (as compared with values at extreme 
resonance), but the SPE peaks are more strongly depressed \cite{JPCM}. Fig. \ref{fig9} 
shows a typical spectrum at $h\nu_i=1642$ meV.

Nevertheless, a modest increase of the peak intensity for $h\nu_i$ well above the band gap may
result not only from large values of $|\langle f|\hat H^+_{e-r}|int\rangle
\langle int|\hat H^-_{e-r}
|i\rangle|^2$ but also from relatively small level broadening (as compared with the
neighboring levels). There could be a set of intermediate states in which $\Gamma_{int}$
takes relatively small values. One can think, for example, of the lowest state in a
sub-band, let us say the $k_z=2$ electron sub-band. Inter sub-band transitions due to
phonon emission are not as fast as intra sub-band transitions \cite{Shah}.
Consequently, the $\Gamma_{int}$ of the lowest state in the sub-band is relatively small.
We will call these states ``excitonic'' states, X. If the product
$\langle f|\hat H^+_{e-r}|X\rangle\langle X|\hat H^-_{e-r}|i\rangle$ is not small, a
peak in the Raman intensity will appear. In the interpretation of Ref. \onlinecite{Danan},
this is an ``incoming'' resonance.

\begin{figure*}[ht]
\begin{center}
\includegraphics[width=.8\linewidth,angle=-90]{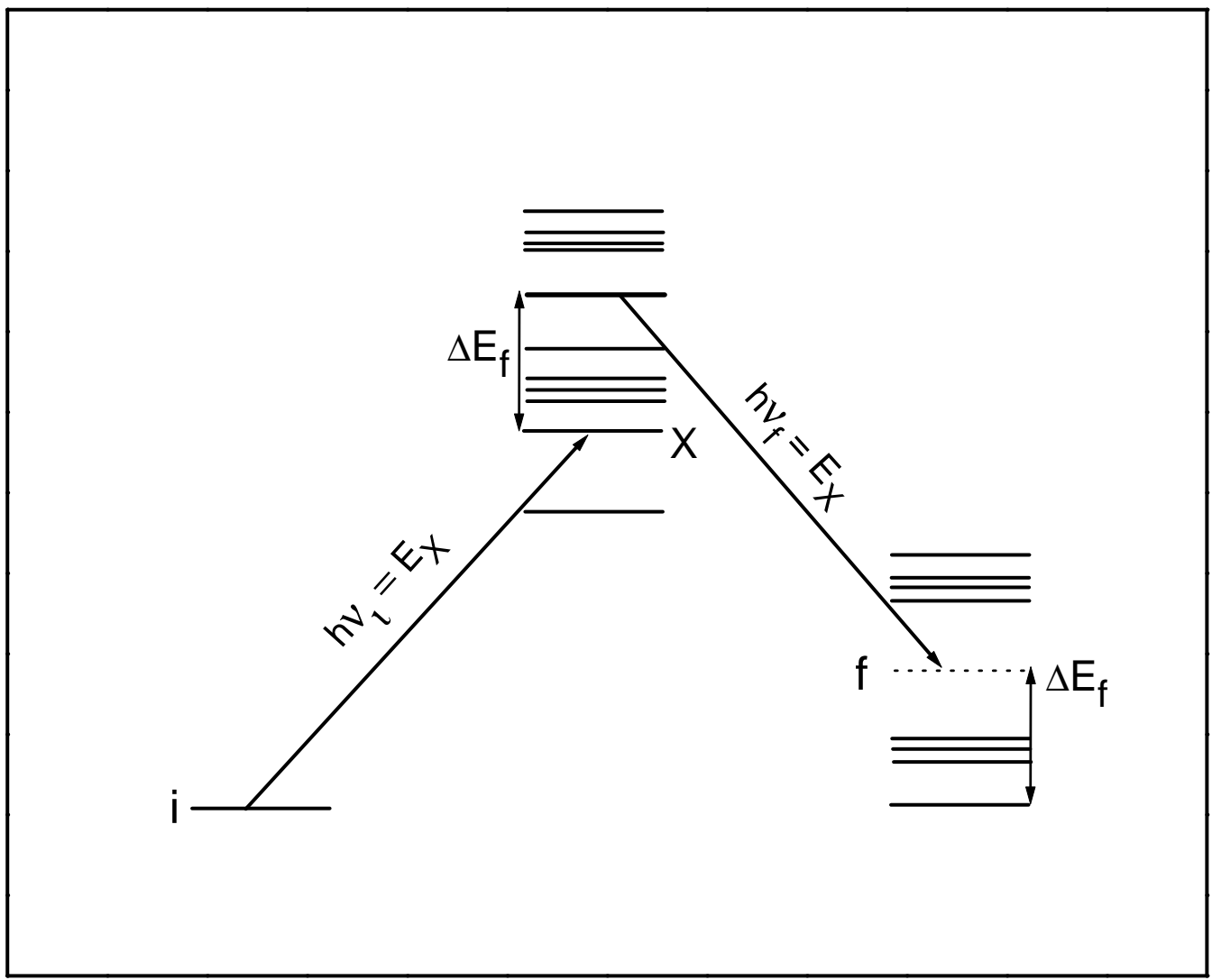}
\caption{Schematic representation of incoming and outgoing Raman
resonances. \label{fig10}}
\end{center}
\end{figure*}

On the other hand, ``outgoing'' resonances correspond to emitted photons with energy
$E_X$, as shown in Fig. \ref{fig10}. The intermediate states are located at excitation
energies $E_X+E_f$. One can think of these states as an exciton on top of a collective
excitation or, conversely, a collective electronic excitation on top of an exciton 
\cite{Lozovik}. The $X$ in this case may correspond to an absorption peak
in the extreme resonance region or well above band gap.

The resonant enhancement of these ``exciton plus plasmon'' states can be due, again, to
big numerators or small denominators in Eq. (\ref{eq1}). Relatively small $\Gamma_{int}$
could be related to the collective nature of these states or to the relative isolation
from neighboring levels. 

\begin{figure*}[ht]
\begin{center}
\includegraphics[width=.7\linewidth,angle=-90]{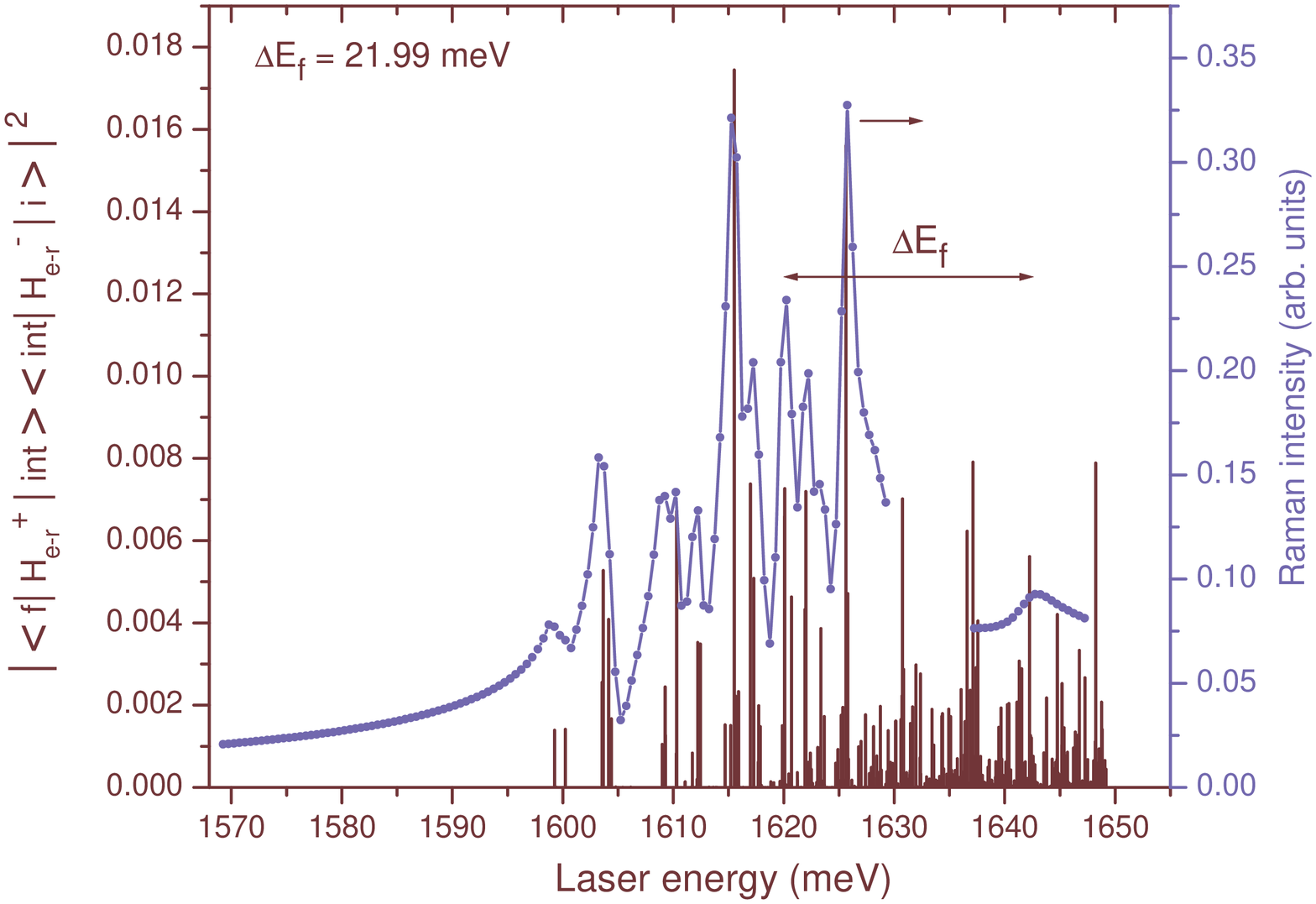}
\caption{(Color online) Intensity of the CDE monopolar Raman peak as a 
function of $h\nu_i$. An outgoing resonance at 1642 meV is modelled. 
\label{fig11}}
\end{center}
\end{figure*}

To illustrate the effect of resonant enhancement of Raman peaks, we show in Fig. 
\ref{fig11} the intensity of the CDE peak as a function of $h\nu_i$. The entire range
of variation is shown for completeness, i.e., below band gap excitation, extreme 
resonance, and well above band gap excitation. In the extreme resonance case, the 
enhancement is related to absorption maxima as mentioned above. On the other hand, 
for $h\nu_i$ well above band gap,
we pick up an intermediate state with energy $E_X+E_{CDE}$, where the
exciton level $X$ corresponds to the absorption maximum at 1620 meV 
(see Fig. \ref{fig8}a), and $E_{CDE}\approx 22$ meV is the energy of the charge monopolar
collective state. For this state, we chose $\Gamma_{int}=2$ meV. The effect is an
enhancement of the intensity around 1642 meV, as shown in the figure, that 
according to Ref. \onlinecite{Danan} is an outgoing resonance.

This calculation reveals that the dependence of $\Gamma_{int}$ on $E_{int}$ may dictate
the qualitative features of Raman scattering with well above band gap excitation. A
consistent treatment in which the $\Gamma_{int}$ are computed for each $|int\rangle$ is
left for future study.

\section{Conclusions}
\label{sec4}

This theoretical investigation of the role of resonance excitation in
electronic Raman scattering from qdots has revealed many hitherto unsuspected
features and details. In general terms, the Raman intensities of the SPEs, 
CDEs, and SDEs are strongly affected when the incident light energy is swept
from below, through, and above the quantum dot band gap. Incoming resonances 
produce a rapid variation in intensity for excitation energies just above
the band gap, and outgoing resonances are predicted for higher excitation 
energies, as observed experimentally. In fact, observation of the Raman
intensity of just one SPE, for example, as a function of the incident light
energy can provide precise details of the optical absorption spectrum and 
density of states. The role of damping in the intermediate states has been 
shown to be a significant factor in determining these resonances and
deserves further theoretical analysis.

Another aspect of this work is the unravelling of the complexity of
features in polarized Raman spectroscopy of qdots. This spectral complexity
in dots with large numbers of electrons has been evident from the first 
experiments. These calculations have shown what excitations dominate in
which polarizations and point the way to a better control of what is
measured in future experiments.

Experimentally, the predicted selective resonances have advantages and 
disadvantages. By employing resonance excitation, a particular final state 
can be enhanced over its companions and thus make it easier to identify.
On the other hand, the rapid variation in intensity of the individual and
numerous SPEs makes it difficult to uniquely identify them. In this regard,
the fine structure evident for SPEs from these calculations should be
explored by performing high resolution Raman spectroscopy in future. The
best situation for evaluating the SPEs would be for an excitation energy
about 30 meV below the band gap, where some overall resonance enhancement
occurs but the contribution from band gap photoluminescence would be
weak. No such experiments have been performed to date.

\begin{acknowledgments}

Part of this work was performed at the Institute for
Microstructural Sciences (IMS), National Research Council, Ottawa. 
A.D. acknowledges the hospitality and support of IMS.

\end{acknowledgments}

\appendix
\section{Matrix elements of the Kohn-Luttinger Hamiltonian}
\label{appendixA}

In the present appendix, we give the matrix elements of the
Kohn-Luttinger Hamiltonian entering Eq. (\ref{eq8}). Results are
presented for the more general case when a magnetic field, $B$, is
applied in the $z$ direction. We use the following set of
parameters \cite{parameters}:

\begin{eqnarray}
\gamma_1=6.790, \quad \gamma_2=1.924, \quad \gamma_3=2.681,
 \nonumber\\ \bar\gamma=(\gamma_2+\gamma_3)/2, \quad \kappa=1.2,
 \quad q=0.04 \nonumber
\end{eqnarray}

The Kohn-Luttinger Hamiltonian has the following structure in the
$m_j$ variable:

\begin{eqnarray}
\cal{H}_{KL}=\left(\begin{array}{cccc}
 H_{3/2} & S & R & 0 \\
 S^{\dagger} & H_{1/2} & 0 & R \\
 R^{\dagger} & 0 & H_{-1/2} & -S \\
 0 & R^{\dagger} & -S^{\dagger} & H_{-3/2}
 \end{array}\right)
\end{eqnarray}

\noindent The $H$ terms are diagonal in oscillator and $k_z$
indexes. They are given by:

\begin{widetext}
\begin{eqnarray}
H_{\pm 3/2}=\hbar\Omega_e \frac{m_e (\gamma_1+\gamma_2)}{m_0}
 (2 k+|l_h|+1)+\frac{\hbar^2}{2 m_0} (\gamma_1-2 \gamma_2)
 \frac{k_z^2\pi^2}{L^2}-\frac{\hbar\omega_{ce}}{2}
 \frac{m_e (\gamma_1+\gamma_2)}{m_0}l_h \pm \mu_B
 \left(3\kappa+\frac{27 q}{4}\right)B,
\end{eqnarray}

\begin{eqnarray}
H_{\pm 1/2}=\hbar\Omega_e \frac{m_e (\gamma_1-\gamma_2)}{m_0}
 (2 k+|l_h|+1) + \frac{\hbar^2}{2 m_0} (\gamma_1+2 \gamma_2)
 \frac{k_z^2\pi^2}{L^2}-\frac{\hbar\omega_{ce}}{2}
 \frac{m_e (\gamma_1-\gamma_2)}{m_0}l_h \pm \mu_B
 \left(\kappa+\frac{q}{4}\right)B,
\end{eqnarray}
\end{widetext}

\noindent where $m_0$ is the electron mass in vacuum, $\mu_B$ is
the atomic Bohr magneton, $\omega_{ce}$ is the electron cyclotron
frequency, $\Omega_e=\sqrt{\omega_0^2+\omega_{ce}^2/4}$, and $k$,
$l_h$ are the radial and angular momentum quantum numbers
corresponding to the 2D oscillator state.

The matrix elements of the $S$ and $R$ operators are written in
the following form:

\begin{widetext}
\begin{eqnarray}
\langle k,l_h,k_z|S|k',l'_h,k'_z\rangle &=&
 \frac{\sqrt{3}\hbar^2\gamma_3}{m_0}\sqrt{\frac{m_e\Omega_e}{\hbar}}
 \frac{4 k_z k'_z}{L (k_z^2-k_z^{'2})} \delta_{l_h,l'_h+1}
 \delta_{k_z+k'_z,odd} \nonumber\\ \nonumber\\
 &\times& \left\{\begin{array}{ll}
 \sqrt{k'+1}\left(1-\omega_{ce}(2\Omega_e) \right)
 \delta_{k,k'+1}+\sqrt{k'+|l'_h|}\left(1+\omega_{ce}/(2\Omega_e)
 \right) \delta_{k,k'}; & l'_h\le -1 \\ & \\
 \sqrt{k'} \left(1+\omega_{ce}/(2\Omega_e)
 \right) \delta_{k,k'-1} +\sqrt{k'+l'_h+1}\left(1-\omega_{ce}/
 (2\Omega_e) \right) \delta_{k,k'}; & l'_h\ge 0,
 \end{array}\right.
\end{eqnarray}

\begin{eqnarray}
\langle k,l_h,k_z|R|k',l'_h,k'_z\rangle &=&
 \frac{\sqrt{3}\bar\gamma m_e}{2 m_0}\hbar\Omega_e
 \delta_{l_h,l'_h+2}\delta_{k_z,k'_z} \nonumber\\
 \nonumber\\
 &\times& \left\{\begin{array}{ll}
 \sqrt{(k'+|l'_h|-1)(k'+|l'_h|)}\left(1+\omega_{ce}/
 (2\Omega_e)\right)^2 \delta_{k,k'} &  \\ & \\
 2\sqrt{(k'+1)(k'+|l'_h|)} \left(1-\left(
 \omega_{ce}/(2\Omega_e)\right)^2\right) \delta_{k,k'+1};
 & l'_h\le -2 \\ & \\
 \sqrt{(k'+2)(k'+1)}\left(1-\omega_{ce}/
 (2\Omega_e)\right)^2 \delta_{k,k'+2} &
 \end{array}\right.\nonumber\\ \nonumber\\
 &\times& \left\{\begin{array}{ll}
 -\sqrt{k'(k'+1)}\left(1+\omega_{ce}/
 (2\Omega_e)\right)^2 \delta_{k,k'-1} &  \\ & \\
 -2 k'\left(1-\left(\omega_{ce}/(2\Omega_e)
 \right)^2\right) \delta_{k,k'}; & l'_h=-1 \\ & \\
 -\sqrt{(k'+2)(k'+1)}\left(1-\omega_{ce}/
 (2\Omega_e)\right)^2 \delta_{k,k'+1} &
 \end{array}\right.\nonumber\\ \nonumber\\
 &\times& \left\{\begin{array}{ll}
 \sqrt{(k'-1)k'}\left(1+\omega_{ce}/
 (2\Omega_e)\right)^2 \delta_{k,k'-2} &  \\ & \\
 2\sqrt{k'(k'+|l'_h|+1)} \left(1-\left(
 \omega_{ce}/(2\Omega_e)\right)^2\right)\delta_{k,k'-1};
 & l'_h\ge 0 \\ & \\
 \sqrt{(k'+l'_h+2)(k'+l'_h+1)}\left(1-\omega_{ce}/
 (2\Omega_e)\right)^2 \delta_{k,k'}. &
 \end{array}\right.
\end{eqnarray}

\end{widetext}

\begin{table*}[h]
\begin{tabular}{|c||c|c|c|c|}
\hline
$\sigma \backslash ~m_j $& 3/2 & 1/2 & -1/2 & -3/2 \\
\hline\hline 1/2 & $\varepsilon _{+i}$ &
$\sqrt{2/3}\;\varepsilon_{zi}$ & $\sqrt{1/3}\;\varepsilon_{-i}$ & 0 \\
\hline -1/2 & 0 & $\sqrt{1/3}\;\varepsilon _{+i}$ & $\sqrt{2/3}\;
 \varepsilon_{zi}$ & $\varepsilon_{-i}$ \\
\hline
\end{tabular}
\caption{The quotient $\vec\varepsilon_i\cdot\vec
p_{\sigma,m_j}/(iP)$, where $P$ is the GaAs band constant.}
\label{tab1}
\end{table*}

\begin{thebibliography}{99}
\bibitem{nota} Strictly speaking, there are two second-order terms
 contributing to the Raman scattering, one like in Eq. (\ref{eq1}),
 and a second term, which has a similar structure but with a
 different energy denominator: $E_i-h\nu_f-E_{int}\approx -2 E_{gap}
 \approx -3000$ meV. \cite{text} Our calculations including both
 terms have shown that the contribution from the second one is
 insignificant in the present case.
\bibitem{nota2} To reproduce the experimental conditions of paper
 [\onlinecite{Heitmann}], a harmonic confinement, $\hbar\omega_0$ of 
 around 6 meV should be taken. However, we choose a higher value, 12 meV,
 in order to be closer to the strong confinement situation holding
 for self-assembled qdots\cite{Hawrylak}, which is the immediate interest 
 of planned experiments.
\end{thebibliography}
\end{document}